\documentstyle[preprint,aps]{revtex}
\begin{document}
\title {Ultra-efficient Cooling in Ferromagnet-Superconductor 
	Microrefrigerators}
\author{Francesco Giazotto\thanks{e-mail: giazotto@nest.sns.it}, 
	Fabio Taddei\thanks{also with: ISI Foundation, I-10133 Torino, Italy, and Dipartimento di Metodologie Fisiche e Chimiche per l'Ingegneria (DMFCI), Universita' di Catania, I-95129 Catania, Italy}, Rosario Fazio and Fabio Beltram}
\address{NEST-INFM, Scuola Normale Superiore, I-56126 Pisa, Italy}

\maketitle

\begin{abstract}
A promising scheme for electron microrefrigeration based on 
ferromagnet-superconductor contacts is presented. 
In this setup, cooling power densities 
up to 600 nW/$\mu$m$^2$ can be achieved leading to electronic 
temperature reductions largely exceeding those obtained with existing 
superconductor-normal metal tunnel contacts. 
Half-metallic CrO$_2$/Al bilayers are indicated as ideal candidates for the implementation of the device.  
\end{abstract}

\pacs{PACS numbers: 73.20.-r, 73.23.-b, 73.40.-c}


Heat transport through metal-superconductor interfaces can be
 applied to microcooling~\cite{martinis,pekola,pekola1}. 
The physical mechanism underlying this electronic cooling is quite simple.
When a normal metal (N) is brought in contact with a superconductor (S),
quasi-particle transport is effective only at energies larger than the S gap 
(${\mathcal E} > \Delta$).  In fact, owing to the existence of the gap 
in the energy spectrum of the superconductor, at a given bias across the SN 
system  only electrons possessing energies higher than the 
gap match available single-particle states and can transfer into the S portion of the junction (see Fig. 1(a)). This selective  transfer of "hot" carriers leads to the lowering of the effective electron temperature of the N electrode, even in the regime when electrons  
are thermally decoupled from the lattice. 
This situation can be experimentally realized with SN {\it tunnel} junctions where transport is dominated by quasi-particle dynamics. 
This unique property of Superconductor-Insulator-Normal 
metal (SIN) contacts was successfully employed for the realization of microcoolers \cite{martinis,pekola}. Intrinsically,
however, SIN devices present large values of the contact resistance 
($R_n$) which hinder carrier transfer and lead to a severe limitation in the
achievable cooling powers \cite{pekola}.

Decreasing $R_n$ by using high-transparency interfaces is not a viable route to increase cooling power. In this case, in fact, another process dominates carrier transfer across the junction. At low $R_n$ values, carrier transfer is made possible by Andreev reflection (AR) even at energies ${\mathcal E}< \Delta$ \cite{andr}.  In AR electrons may coherently evolve into holes that retrace the incoming electron 
time-reversed path and transfer a Cooper pair in S (see Fig.\,1(b)). This does lead to increased conductivity, but does not contribute to 
thermal transport through the system. In practical devices a compromise must be reached and an optimal transparency value can be determined for a given microrefrigerator structure~\cite{averin}.

In this Letter we propose a  cooling mechanism alternative to  
traditional SIN junctions which will be shown to be highly efficient. 
Our proposal is based on the use of a thin ferromagnetic  layer (F) 
in good electric contact with S as shown in Fig.1(c). The 
physical basis of the FS microcooler operation  resides 
in the spin-band splitting characteristic of F.
In fact the electron ({\it e}) involved in AR and its 
phase-matched hole ({\it h}) must belong to opposite spin bands (Fig.1(d)); thus, 
suppression of the Andreev current~\cite{beenakker} occurs in an FS junction and
its intensity depends on the degree of the F-layer spin polarization $\mathcal P$. 
In the limit of large $\mathcal P$ and good metallic contact between the F and S electrodes, we observe a drastic suppression of the subgap current, while keeping efficient hot-carrier transfer leading to 
a  considerable thermal flux ${\mathcal J}$. 
 Cooling power densities 
are almost two orders of magnitude larger  
than those of optimized SIN junctions.  
This efficiency enhancement translates in a 
dramatic reduction of the final achievable electron temperature. 

The device is  sketched in Fig.\,1(c). It consists of a NFS microcooler 
biased at a voltage $V$. Since the refrigeration stems from 
the FS portion, we shall focus on this junction only.
Let us consider a ballistic FS contact with a 
perfectly transparent interface ~\cite{barriera}. 
In such a junction
the overall cooling power ${\mathcal J}$ is equal to the sum of terms of
the form \cite{averin}
\begin{equation}
\frac{1}{h}\int d{\mathcal E} \left\{
{\mathcal E} \left[1-B-A\right] -eV\left[1-B+A\right] \right\} \left( f_F -f_S\right)
\label{cool}
\end{equation}
each relative to a given angle of incidence and spin species.
In (\ref{cool}), $f_{F,S}({\mathcal E})$ is the electron Fermi distribution
function in F(S), whereas 
$A({\mathcal E})$ and $B({\mathcal E})$ are the Andreev and 
normal reflection probabilities obtained, in close analogy to
Refs.\cite{beenakker,btk}, as solutions of the
Bogoliubov-de Gennes equations.
The S pairing potential is approximated as a step-function \cite{btk}, 
while F is
described with the usual Stoner model~\cite{beenakker}.
Differences in the effective-mass and  Fermi-velocity 
values between F and S are neglected.
Note that ${\mathcal J}$ is a function of spin
polarization (via the scattering probabilities), bias voltage and temperature.

In order to evaluate the final electron temperature
we need to consider also the mechanisms that can transfer energy into the 
ferromagnet. 
The main contribution is due to electron-phonon coupling and is given by \cite{kautz,urbina} 
${\mathcal J}_{l}=\Sigma {\mathcal V}(T_e^5 -T_l^5)$, 
where ${\mathcal V}$ is the volume of the F electrode, $T_{e(l)}$ is 
the electron (lattice) temperature and $\Sigma$ 
is a material-dependent parameter of the order $10^{-9}$ WK$^{-5}$
$\mu$m$^{-3}$. For our 
purposes we can neglect the power dissipated in the resistance of the  F 
layer~\cite{martinis,pekola}. 
The final electron temperature $T_e$ is determined by the energy-balance
equation
${\mathcal J}({\mathcal P},V,T_e,T_l)+{\mathcal J}_{l}(T_e , T_l)=0$. In writing this expression we set the superconductor temperature equal to the bath temperature $T_l$. 
Note that this is an idealized assumption since the  presence of hot quasi-particles in S, that may strongly decrease the real cooling capability of the device, is expected. In real applications, however, one can overcome this problem through the exploitation of ``quasi-particle traps'' attached to the superconductor itself as described in Ref. \cite{pekola1}.  

Figure\,2(a) shows $T_e$ as a function of bias for 
two starting lattice temperatures (i.e., the system temperatures at $V=0$) $T_l=200$ mK (solid lines) and $T_l=300$ mK (dotted lines).
The calculation 
was performed for half-metallic (${\mathcal P} =1$) FS (thick lines) and for SIN (thin lines)
microcoolers, assuming Al as superconductor 
($\Delta_{Al}= 180 \,\,\mu eV$). Other junctions parameters were taken according to experimental values \cite{martinis,pekola,calcolo}. In particular for the 
SIN-junction specific contact resistance we assumed a value  $0.4$ k$\Omega$\,$\mu$m$^2$, corresponding to very high-quality Cu/Al$_2$O$_3$/Al junctions~\cite{martinis,pekola}.
Figure 2(a) shows the remarkable  $T_e$ reduction provided by  the 
FS cooler with respect to the NIS cooler. 
What shown in Fig. 2(a) is not meant to be a comparison with the best optimized SIN cooler (i.e., a SINIS refrigerator)\cite{pekola}, however it helps in establishing the main differences between the single FS cooling element and the single SIN one.
Even starting from 200 mK, for which the tunnel junction provides its largest cooling effect (a temperature reduction of about 10\% at $eV\simeq \Delta$), the FS cooler
yields $T_e$ of the order of 
10 mK (a temperature reduction of about 95\%). This marked difference stems from the high contact resistance 
of the SIN junction that strongly affects its performance and must be compared
to specific contact resistances as low as $10^{-3}\,\,\Omega \mu$m$^2$ that are currently achieved in highly transmissive FS junctions~\cite{burhman}.
For $eV<\Delta$, the FS-junction cooling power is large and, contrary to 
the SIN device, not localized only at $eV\approx \Delta$. For $eV>\Delta$, 
instead, the increase in the electron temperature is quite rapid owing to the 
high current driven through the FS contact. 

The large cooling-power surface density ${\mathcal J}_{\mathcal A}$ 
makes the FS microrefrigerator a high performance device. Figure\,2(b) shows 
${\mathcal J}_{\mathcal A}(\mathcal P)$ calculated at 
each optimal bias voltage, for some values of 
the lattice temperature. In the inset, the same quantity is plotted for a SIN 
junction versus interface transmission probability ${\mathcal D}$~\cite{averin}. 
The comparison shows that for ${\mathcal P}=1$ at $T\simeq 0.4\,\Delta /k_B$  
it is possible to achieve power densities up to 600 nW/$\mu$m$^2$, i.e. 30 times larger than those for 
the SIN junction at the optimized  transmissivity 
(${\mathcal D}\approx 3\times 10^{-2}$ at $T\simeq 0.3\,\Delta /k_B$). 
In real applications, however, SIN interface transmissivity is 
much smaller (${\mathcal D}\approx 10^{-6}\div 10^{-5}$) \cite{martinis,pekola}, limiting the achievable cooling power density to some pW/$\mu$m$^2$.

The proposed cooling mechanism is robust against various possible sources 
that can lower quasi-particle transmission or enhance AR.
We have evaluated the impact of incomplete polarization of the ferromagnet (${\mathcal P}<1$), of low transmittance 
FS contacts and of spin-flip processes.
Figure\,3 shows the calculated dimensionless cooling power ${\mathcal J}(V)$ for some ${\mathcal P}$ values at  $T=0.4 \,\Delta /k_B$.
For each value of polarization there exists an optimal bias voltage which maximizes ${\mathcal J}$. 
For ${\mathcal P} = 1$, ${\mathcal J}(V)$ is maximized around 
$V\simeq \Delta/e$. Notably, even for ${\mathcal P}= 94\%$ 
there still is a positive heat current.
The inset of Fig.\,3 shows ${\mathcal J}(T)$ calculated at the optimum bias 
voltage for some values of ${\mathcal P}$. The heat current is maximum 
for $T\simeq 0.4\, \Delta/k_B$. 
In addition, lowering ${\mathcal P}$ 
leads to a reduction of the useful temperature window for electron cooling.
The same behavior was verified in the presence of spin-flip scattering events. We calculated this by allowing a random misalignment of magnetic moments within a given
angle from the perfect alignment, which produces a decrease in the current
polarization. (There is, however, no evidence that such processes occur in CrO$_2$/S nano-junctions
which are the system we suggest for the implementation of the FS microcooler.)
An important figure of merit of the refrigerator is represented by its coefficient of performance (COP),  which is  the ratio between the cooling power and the total input power. Our calculations show that for ${\mathcal P}= 1$ in the $0.2\div 0.4\, \Delta/k_B$ temperature range the COP, evaluated at the optimum bias voltage, exceeds  20$\%$ reaching its highest value (23$\%$) at about $T\simeq 0.3\, \Delta/k_B$. For ${\mathcal P}= 96\%$  we obtain about $10\%$ efficiency at the same temperature.

In order to simulate a non-perfectly transparent FS contact caused by 
fabrication procedures or mismatch in material parameters~\cite{blonder}, 
we have inserted an insulating barrier at the interface. The numerical 
simulation reveals that for ${\mathcal P}=1$ and low transmittance interface (e.g., ${\mathcal D}=0.1$), the variation in the final electron temperature is
not appreciable.

The FS junction discussed so far is part of the NFS device of Fig.\,1(c). We estimated the impact of the F-layer thickness on the NFS-device cooling power.
We verified that cooling effects comparable to the FS
case can be reached for thickness values of the order of a few nm (i.e. corresponding to the magnetic length $\hbar v_F/h_0$, $h_0$ being the
F exchange field and $v_F$ the Fermi velocity).

In conclusion we should like to indicate half-metallic CrO$_2$ (chromium dioxide)~\cite{gambino,hm} in combination with Al as the ideal candidate for the implementation 
of these devices. The well-developed CrO$_2$ technology~\cite{CrO2} can in fact be used to realize complex FS arrays coupled with normal electrodes for optimal 
device geometries.

We acknowledge a useful correspondence with J. P.  Pekola.
This work was supported by INFM under the PAIS projects EISS and TIN.



\begin{figure}
\caption{ (a) Biasing the SIN junction around the energy gap $\Delta$ allows more energetic electrons {\it e} to tunnel into S, thus {\it cooling} the N electron population (see text). (b) Schematic description of Andreev reflection  at a NS contact. (c) Scheme of the proposed NFS microrefrigerator. The NF junction is supposed to be a highly transmissive electric contact. (d) Schematic representation of the principle of operation of the NFS microcooler. For ${\mathcal P}=1$, AR is hindered by the absence of available states for reflected holes,  {\it h}. This subgap electron-transport suppression mechanism allows the operation of the microrefrigerator in the presence of efficient carrier transfer to S.}
\label{F1}
\end{figure}

\begin{figure}
\caption{ (a) Electron temperature $T_e$ versus bias voltage for a half-metallic (${\mathcal P} =1$) FS (thick lines) and a SIN (thin lines) microcoolers for two starting bath temperatures $T_l$ at $V=0$: 300 mK (dotted curves) and 200 mK (solid lines).
(b) Maximum cooling-power surface density $\mathcal J_{\mathcal A}$ versus spin polarization ${\mathcal P}$ for various temperatures. The inset shows the same quantity for a SIN junction as a function of  contact transmissivity ${\mathcal D}$.
}
\label{F2}
\end{figure}

\begin{figure}
\caption{ Dimensionless cooling power ${\mathcal J}$ of a FS contact versus bias voltage at $T=0.4\,\,\Delta /k_B$ for several ${\mathcal P}$ values. The inset shows the heat current calculated at the optimal bias voltage versus temperature for some ${\mathcal P}$ values.
}
\label{F3}

\end{figure}



\begin{references}
\bibitem{calcolo} 
	We assumed a junction area of 0.1 $\mu$m$^2$, an active 
	volume of the FN side 
	${\mathcal V} =0.5$ $\mu$m$^3$, $\Sigma =2\times 10^{-9}$ WK
	$^{-5}$$\mu$m$^{-3}$ (typical of Cu), 
	and $R_{n}^{FS}=28$ m$\Omega$ (assuming a Fermi energy ${\mathcal E}_{F}=5\,\,eV$).
\end{references}
\end{document}